\documentclass[english]{tlp}

\usepackage[latin1]{inputenc}
\usepackage{babel}
\usepackage{graphics}
\usepackage{graphicx}
\usepackage{aopmath}
\usepackage{amsfonts}
\usepackage{amssymb}
\usepackage{color}
\usepackage{multirow}

\newtheorem{example}{Example}       
\newtheorem{definition}{Definition} 

\newcommand{\tts}{\tt \small}

\newcommand{\aaL}{\langle{\hspace*{-2.5pt}}\langle}
\newcommand{\aaR}{\rangle{\hspace*{-2.5pt}}\rangle}

\newcommand{\eop}{\hfill$\Box$}
\usepackage{color}

\newcommand{\exm}

\title{Proving Correctness of Imperative Programs by Linearizing Constrained Horn Clauses}

\author[E. De~Angelis, F. Fioravanti, A. Pettorossi, M. Proietti]
{EMANUELE DE ANGELIS, FABIO FIORAVANTI\\
DEC, University `G. d'Annunzio', Pescara, Italy\\
\email{\{emanuele.deangelis,fabio.fioravanti\}@unich.it}
\and
ALBERTO PETTOROSSI\\
DICII, Universit\`a di Roma Tor Vergata, Roma, Italy\\
\email{pettorossi@disp.uniroma2.it}
\and
MAURIZIO PROIETTI\\
CNR-IASI, Roma, Italy\\
\email{maurizio.proietti@iasi.cnr.it} 
}

\begin{document}

\pagestyle{plain}

\maketitle

\begin{abstract}
We present a method for verifying the correctness of 
imperative programs which is based on the automated transformation 
of their specifications.
Given a program \textit{prog}, we consider a partial correctness 
specification of the form
$\{\varphi\}\, \textit{prog}\, \{\psi\}$,
where the assertions $\varphi $ and~$\psi$ are predicates
defined by a set~\textit{Spec} of possibly recursive 
Horn clauses with linear arithmetic ({\it LA})
constraints in their premise (also called {\it constrained Horn clauses}).
The verification method consists in
constructing  a set \textit{PC} of constrained Horn clauses
whose satisfiability implies that $\{\varphi\}\, \textit{prog}\, 
\{\psi\}$ is valid.
We highlight some limitations of state-of-the-art constrained Horn clause solving
methods, here called {\it{LA-solving methods}}, 
which prove the satisfiability of the clauses  by looking for linear arithmetic interpretations of the 
predicates. 
In particular, we prove that there exist some specifications 
that cannot be 
proved valid by any of those {\it LA}-solving methods. 
These specifications require the proof
of satisfiability of a set \textit{PC} of constrained Horn clauses that contain
{\it nonlinear clauses} (that is, clauses with more than one atom in their premise).
Then, we present a transformation, called {\it linearization}, that 
converts \textit{PC} into
a set of {\it linear} clauses (that is, clauses with at most one 
atom in their premise).
We show that several specifications 
that could not be proved valid by  {\it LA}-solving  methods,
can be proved valid after linearization.
We also present a strategy for performing linearization in an automatic way and
we report on some experimental results obtained by using a preliminary implementation
of our method.

\medskip

\noindent
{\em To appear in Theory and Practice of Logic Programming (TPLP), Proceedings of ICLP 2015.}

\end{abstract}

\begin{keywords}
Program verification,
Partial  correctness specifications,
Horn clauses,
Constraint Logic Programming,
Program transformation.\vspace{-2mm}
\end{keywords}

\section{Introduction}
\label{sec:intro}

One of the most established methodologies for specifying
and proving the correctness of imperative programs is based on the 
Floyd-Hoare axiomatic approach
(see~\cite{Hoa69}, and also~\cite{Ap&09} for a recent presentation dealing with 
both sequential and concurrent programs). 
By following this approach, the {\it partial correctness} of a program 
\textit{prog} is formalized by a triple
$\{\varphi\}\, \textit{prog}\, \{\psi\}$, also called {\it partial 
correctness specification}, 
where the {\it precondition} $\varphi$ and the
{\it postcondition} $\psi$ are assertions in
first order logic, meaning that
if the input values of \textit{prog} satisfy $\varphi$
and program execution terminates, then the output values satisfy~$\psi$.

It is well-known that the problem of checking
partial correctness of programs with respect to given 
preconditions and postconditions
is undecidable. 
In particular, the undecidability of partial correctness is due to the fact
that in order to prove in Hoare logic the validity of a 
triple $\{\varphi\}\ \textit{prog}\ \{\psi\}$, 
one has to look for suitable auxiliary assertions, 
the so-called {\it invariants}, in an infinite space of formulas, and also
to cope with the undecidability of logical consequence.

Thus, the best way of addressing the problem of the automatic verification of
programs is to design {\it incomplete} methods, that is, methods
based on restrictions of first order logic, which work well in the practical cases of interest.
To achieve this goal, some methods proposed in the literature
in recent years use 
{\it linear arithmetic constraints} as the assertion language and {\it constrained 
Horn clauses} as the formalism to express and reason about program 
correctness~\cite{Bj&12,De&14c,Gr&12,Ja&12,Pe&98,PoR07,Ru&13}.

Constrained Horn clauses are clauses with {\it at most one} atom in their conclusion and a conjunction of atoms
and constraints over a given domain in their premise. In this paper 
we will only consider constrained Horn clauses with linear arithmetic constraints.
The use of  this formalism has the advantage that logical consequence 
for linear arithmetic constraints is decidable and, moreover, 
reasoning within constrained Horn clauses is supported by very effective
automated tools, such as {\it Satisfiability Modulo Theories} (SMT)
solvers~\cite{DeB08,MaS13,Ru&13} and
{\it constraint logic programming} (CLP) inference systems~\cite{JaM94}.
However, current approaches to correctness proofs based on constrained Horn clauses 
have the disadvantage that they only consider specifications whose preconditions and  
postconditions are linear arithmetic constraints.

In this paper we overcome this limitation and propose an approach 
to proving general specifications of the form 
$\{\varphi\}\, \textit{prog}\, \{\psi\}$, where $\varphi$ 
and~$\psi$ are predicates defined by a set of possibly recursive constrained Horn clauses 
(not simply linear arithmetic constraints), 
and {\it prog} is a program written in a C-like imperative language.

First, we indicate how to construct 
a set \textit{PC}  of constrained Horn clauses (\textit{PC} stands for 
partial correctness), 
starting from: (i)~the assertions $\varphi $ and $\psi$, 
(ii)~the program~\textit{prog}, and (iii)~the 
definition of the operational semantics of the language in which
 \textit{prog} is written, 
such that, if $\textit{PC}$ is satisfiable, then the partial correctness 
specification $\{\varphi\}\, \textit{prog}\, \{\psi\}$ is valid.

Then, we formally show that there are sets \textit{PC} of constrained Horn clauses
encoding partial correctness specifications, whose satisfiability cannot  be
proved by current methods,
here collectively called {\it \mbox{{\textit {LA}}-solving} methods}
({\it LA} stands for linear arithmetic).
This limitation is due to the fact that 
\mbox{{\textit {LA}}-solving} methods try to prove satisfiability
by interpreting the predicates as linear arithmetic constraints.

For these problematic specifications, the set~\textit{PC} of constrained Horn clauses contains
{\it nonlinear} clauses, that is, clauses with more than one atom in their premise.

Next, we present a transformation, which we 
call {\it linearization}, that 
converts the set \textit{PC} into
a set of {\it linear} clauses, that is, clauses with at most one 
atom in their premise.
We show that linearization preserves satisfiability and also 
increases the power of {\textit {LA}}-solving, in the sense that
several specifications that could not be proved valid by {\textit {LA}}-solving methods,
can be proved valid after linearization.
Thus, linearization followed by {\textit {LA}}-solving is 
strictly more powerful than {\textit {LA}}-solving alone.

The paper is organized as follows. In Section~\ref{sec:encoding}
we show how a class of partial correctness specifications can be translated into
constrained Horn clauses.
In Section~\ref{sec:Limitations} we prove that {\textit {LA}}-solving methods
are inherently incomplete for proving the satisfiability of constrained Horn clauses.
In Section~\ref{sec:transform} we present a strategy for automatically performing 
the linearization transformation,
we prove that it preserves {\textit {LA}}-solvability, and (in some cases) it is able to
transform constrained Horn clauses that are not {\textit {LA}}-solvable into
constrained Horn clauses that are {\textit {LA}}-solvable. 
Finally, in Section~\ref{sec:experiments}
we report on some preliminary experimental results obtained by using
a proof-of-concept implementation of the method.

\section{Translating Partial Correctness into Constrained Horn Clauses}
\label{sec:encoding}

We consider a C-like imperative programming language with integer 
variables, assignments, conditionals, while loops, and goto's.
An imperative  program is a sequence of labeled commands (or commands, for short), 
and in each program there is a unique
$\mathtt{halt}$ command that, when executed, causes program termination.

The semantics of our language is defined by a {\it transition relation},  denoted 
$\Longrightarrow$, between {\it configurations}. Each configuration is a pair
$\aaL \ell\!:\!c, \delta\aaR$ of a labeled command~$\ell\!:\!c$ and an 
{\it environment}~$\delta$.
An {environment}~$\delta$ is a function 
that maps every integer variable identifier $x$ to its value $v$ 
in the integers~$\mathbb Z$.
The definition of the relation $\Longrightarrow$ is similar to that of the
`{small step}' operational semantics presented in~\cite{Rey98},
and is omitted.
Given\,a\,program \textit{prog},\,we\,denote\,by 
$\ell_0\!:\!c_0$ its\,first\,labeled command.

We assume that all program executions are {\it deterministic} in the sense that,
for every environment~$\delta_0$,
there exists a unique, maximal (possibly infinite) 
sequence of configurations, called {\it computation sequence}, 
of the form: $\aaL \ell_0\!:\!c_{0}, \ \delta_0\aaR \Longrightarrow$
\mbox{$\aaL \ell_1\!:\!c_{1}, \ \delta_1\aaR$} $ \Longrightarrow \cdots $.
We also assume that every {\it finite} computation sequence ends in the configuration
$\aaL \ell_{h}\!:\!\mathtt{halt}, \ \delta_n\aaR$, for some environment $\delta_n$.
We say that  a program \textit{prog}  {\it terminates} for~$\delta_0$ iff 
the computation sequence
starting from the initial configuration $\aaL \ell_0\!:\!c_{0}, \ \delta_0\aaR$ 
is finite.

\newpage

\subsection{Specifying Program Correctness}
\label{subsec:Spec}

First we need the following notions about constraints, 
constraint logic programming, and constrained Horn clauses. 
For related notions with which the reader is not
 familiar, he may refer to~\cite{JaM94,Llo87}.

A \textit{constraint}
is a linear arithmetic equality (=) or inequality ($>$)
over the integers~$\mathbb Z$, or a conjunction
or a disjunction of constraints. For example, 
$2.\!X\!\geq\!3.\!Y\! - 4$ is a constraint.
We feel free to say
`linear arithmetic constraint', instead of `constraint'.
We denote by~$\mathcal C_{{LA}}$ the set of all constraints.
An {\it atom} is an atomic formula of the form $p(t_{1},\ldots,t_{m})$,
where $p$ is a predicate symbol not in $\{=,>\}$ and 
$\mathit{t_{1},\ldots,t_{m}}$ are terms.
Let {\it Atom} be the set of all atoms.
A~{\it definite clause} is an implication of the form  
$A\leftarrow c, G$, where in the conclusion (or {\it head\/}) $A$ is an atom, 
and in the
premise (or {\it body\/}) $c$ is a constraint, and~$G$ is a (possibly empty)
conjunction of atoms.
A~{\it constrained goal} (or simply, a {\it goal}\/) is  an implication of the form  
$\textit{false} \leftarrow c, G$.
A~{\it constrained Horn clause} (CHC) (or simply, a {\it clause}) 
is either a definite 
clause or a constrained goal. 
A {\it constraint logic program} (or simply, a CLP {\it program})  
is a set of definite clauses.
A~clause {\it over the integers} is a clause that has no function 
symbols except for integer constants, addition, and
multiplication by integer constants.

The semantics of a constraint $c$ 
is defined in terms of the usual interpretation, denoted by ${\textit{LA}}$,
over the integers $\mathbb Z$. We write ${\textit{LA}}\models c$ to denote
that $c$ is true in~${\textit{LA}}$.
Given a set $S$ of constrained Horn clauses, an
{\it LA-interpretation} is an interpretation for the language of
$S$ that agrees with {\it LA}
on the language of the constraints.
An {\it LA-model} of  $S$
is an {\it LA-interpretation} that makes all clauses of $S$ true.
A set of constrained Horn clauses is {\it satisfiable} if it has an {\it LA-model}.
A CLP program $P$ is always satisfiable and has a 
{\it least ${\textit{LA}}$-model}, denoted $M(P)$. 
We have that a set $S$ of constrained Horn clauses is {\it satisfiable} iff 
$S\!=\! P \!\cup\! G$, where~$P$ is a CLP program, $G$ is a set  of goals, and
$M(P)\models G$.
Given a first order formula $\varphi$, we denote by $\exists (\varphi)$
its {\it existential closure} and by $\forall (\varphi)$ its 
{\it universal closure}.

Throughout the paper we will consider partial correctness 
specifications which are particular triples of the form
\mbox{$\{\varphi\} $ {\textit{prog}} $
\{\psi\}$} defined as follows.
\vspace{-1mm}

\begin{definition}[{{Functional Horn Specification}}]\label{def:spec} 
{\rm 
A partial correctness triple
\mbox{$\{\varphi\} $ {\textit{prog}} $
\{\psi\}$} is said to be a {\it functional Horn specification} if the following 
assumptions hold, where the predicates $\textit{pre}$ and $f$ are 
assumed to be defined by a CLP program \textit{Spec\/}:

\noindent\hangindent=4mm
(1) $\varphi$ is the formula: 
$z_{1}\!=\!p_1\wedge\ldots \wedge z_{s}\!=\!p_s 
\wedge \textit{pre\/}(p_1,\ldots,p_s)$,
where $z_{1},\ldots, z_{s}$ are the variables occurring in \textit{prog}, and
 $p_1,\ldots,p_s$ are variables (distinct from the~$z_{i}$'s), called
{\it parameters} (informally,  $\textit{pre}$ determines the 
initial values of the $z_{i}$'s); 

\noindent\hangindent=4mm
(2)  $\psi$ is the atom $f(p_1,\ldots,p_s,z_k)$, 
where $z_{k}$ is a variable in $\{z_{1},\ldots, z_{s}\}$ (informally, 
$z_{k}$ is the variable whose final value is the result of 
the computation of~\textit{prog});

\noindent\hangindent=4mm
(3) \textit{f}\/ is a relation which is  {\it total on pre\/} and {\it functional}, in the sense that 
the following two properties hold (informally, $f$ is the function
computed by \textit{prog}):

(3.1)\,$M(\textit{Spec})\! \models \forall p_1,\ldots,p_s$.
$\textit{pre\/}(p_1,\ldots,p_s) 
\rightarrow \exists y$. $f(p_1,\ldots,p_s,y) $\nopagebreak

(3.2)\,$M(\textit{Spec})\! \models 
\forall p_1,\ldots,p_s,y_1,y_2$. $f(p_1,\ldots,p_s,y_1) 
\wedge f(p_1,\ldots,p_s,y_2)\! \rightarrow \!y_1\!=\!y_2$.~\eop
}\vspace*{-1mm}
\end{definition}

\noindent
We say that a functional Horn 
specification \mbox{$\{\varphi\} $ {\textit{prog}} $
\{\psi\}$} 
is {\it valid\/}, or \textit{prog} is 
partially correct with respect to $\varphi$ and $\psi$, iff
for all environments
$\delta_{{0}}$ and $\delta_{n}$,

\noindent 
{\it if} $M(\textit{Spec})\!\models \textit{pre}(\delta_{{0}}(z_{1}\!),
\ldots,
\delta_{{0}}(z_{s}))$ holds (in words,  $\delta_0$ satisfies \textit{pre}) 
and $\aaL \ell_{0}\!:\!c_{0}, 
 \delta_{{0}}\aaR$ $\Longrightarrow ^*$
\mbox{$\aaL \ell_{h}\!:\!{\mathtt{halt}},  \delta_{n}\aaR$} holds (in words, 
\textit{prog} terminates for $\delta_0$) holds, 
{\it then} $M(\textit{Spec})\models f(\delta_{{0}}(z_{1}),\ldots,\delta_{{0}}(z_{s}),\delta_{n}(z_k))$ holds (in words, $\delta_{n}$ satisfies the postcondition).

The relation $r_{\textit{\small prog}}$
computed by \textit{prog} according to the operational semantics of 
the imperative language, is defined by the CLP 
program~\textit{OpSem} made out of: (i)~the following clause~$R$
(where, as usual, variables are denoted by upper-case letters):

\noindent
$R$.~~~ $r_{\textit{\small prog}}(\!P_1,\ldots,\!P_s,\!Z_k) 
\leftarrow \textit{initCf\/}
(\!C_0,P_1,\ldots,\!P_s),\; \textit{reach}(\!C_0,\textit{C}_h),\; \textit{finalCf\/}(\!C_h,Z_k)$

\noindent
where:

\noindent\hangindent=5mm
(i.1)~$\textit{initCf\/}(\!C_0,\!P_1,\!\ldots,\!P_{\!s})$
represents the initial configuration $\textit{C}_0$, where the variables
$z_1,\!\ldots\!,\!z_s$ are bound to the values $P_1,\!\ldots\!,\!P_s$, respectively,
and $\textit{pre}(\!P_1,\!\ldots\!,\!P_s)$ holds,

\noindent\hangindent=5mm
(i.2)~$\textit{reach\/}(\!C_0,C_h)$ represents the transitive closure
$\Longrightarrow ^*$ of the transition relation $\Longrightarrow$, 
which in turn is represented by a predicate $\textit{tr\/}(\!C_1,C_2)$ 
that encodes the operational semantics, that is, the
{\it interpreter} of our imperative language,
by relating a source configuration $C_{1}$ to a target configuration $C_{2}$,

\noindent\hangindent=5mm
(i.3)~$\textit{finalCf\/}(\!C_h,Z_k)$ represents the final configuration 
$\textit{C}_h$, where the variable $z_k$ is bound to the value $Z_k$,

\noindent
and (ii)~the clauses for the predicates 
$\textit{pre\/}(P_1,\ldots,P_s)$ and 
$\textit{tr\/}(\!C_1,\!C_2)$. The 
clauses for the predicate $\textit{tr\/}(\!C_{1},\!C_{2})$ are defined 
as indicated in~\cite{De&14c},
and are omitted for reasons of space.

\vspace{-1mm}
\begin{example}[Fibonacci Numbers]\label{ex:fib}
\begin{rm}
Let us consider the following program \textit{fibonacci}, that returns as value of
{\tt u} 
the {\tt n}-th Fibonacci number, for any {\tt n} $(\geq {\mathtt 0})$, having 
initialized {\tt u} 
to {\tt 1} and {\tt v} to {\tt 0}.
\vspace{.5mm}

\noindent
$\left[
\begin{array}{llr}
{\texttt{0: while\ (n>0)\ \{\ t=u;\  u=u+v;\  v=t; \ n=n-1\ \}}}& & 
\hspace{23mm} ${\textit{fibonacci}}$\\
{\texttt{h: halt}}\\
\end{array}
\right.
$

\vspace{.5mm}
\noindent
The following is a functional Horn specification of the partial correctness of the program \textit{fibonacci}:

\hspace{10mm}{\rm{\{}}{\tt n=N,\,N>=0,\,u=1,\,v=0,\,t=0}{\rm{\}}} ~~~~{\textit{fibonacci}}~~~~
{\rm{\{}}{\tt fib(N,u)}{\rm{\}}}\hfill $(\ddagger)$\nopagebreak

\noindent
where {\tt N} is a parameter and 
{\tt fib} is defined by the following CLP program:

\vspace{.5mm}
\noindent

\noindent
${\scriptstyle{\left[
\begin{array}{llr}
{\small\texttt{S1.~fib(0,1).}}& & \hspace{-24mm}
 {\normalsize{\textit{Spec}_{\textit{\small fibonacci}}}}\\
{\small\texttt{S2.~fib(1,1).}}\\
{\small \texttt{S3.~fib(N3,F3)\,:- 
N1>=0,\,N2=N1+1,\,N3=N2+1,\,F3=F1+F2,\,fib(N1,F1),\,fib(N2,F2).}}\\
\end{array}
\right.
}}$

\vspace{.5mm}
\noindent

\noindent
For reasons of conciseness, in the above specification $(\ddagger)$ we have 
slightly deviated from Definition~\ref{def:spec}. In particular, 
we did not introduce the predicate symbol {\it pre}, and 
in the precondition and postcondition 
we did not introduce the parameters which have constant values.

The relation {\tt r\_fibonacci} computed by the program \textit{fibonacci} 
according to the operational semantics,
is defined by the following CLP program:

\noindent
${\scriptstyle{\left[
\begin{array}{lll}
 & & \hspace*{-27mm}\raisebox{1mm}{$\textit{OpSem}_{\textit{\small fibonacci}}$}\\[-1.5mm]
{\small\texttt{R1.~r\_fibonacci(N,U)\,:- initCf(C0,N),\,reach(C0,Ch),\,finalCf(Ch,U).}}& & \\
{\small\texttt{R2.~initCf(cf(LC,E),N) :- N>=0, U=1, V=0, T=0, firstCmd(LC),}}& & \\
\hspace{28mm}{\small\texttt{env((n,N),E), env((u,U),E), env((v,V),E), env((t,T),E).}}& & \\
{\small\texttt{R3.~finalCf(cf(LC,E),U) :- haltCmd(LC), env((u,U),E).}}\\
\end{array}
\right.
}}$

\smallskip
\noindent
where: (i)~{\tt firstCmd(LC)} holds for the command
with label {\tt 0} of the  program \textit{fibo\-nacci};
(ii)~{\tt env((x,X),E)}  holds iff in the environment {\tt E} the variable 
{\tt x} is bound to the value of {\tt X};
(iii)~in the initial configuration {\tt C0} the environment {\tt E}
binds the variables {\tt n}, {\tt u}, {\tt v}, {\tt t} 
to the values
{\tt N} ({\tt >=0}), {\tt 1}, {\tt 0}, and {\tt 0}, respectively; and
(iv)~{\tt haltCmd(LC)} holds for the labeled command {\tt h:\,halt}.\eop
\end{rm}
\end{example}

\subsection{Encoding Specifications into Constrained Horn Clauses}
\label{subsec:PC}

In this section we present the encoding of the 
validity problem of functional Horn specifications into the satisfiability 
problem of CHC's.
 
For reasons of simplicity we assume that in \textit{Spec} 
no predicate depends on $f$ (possibly, except for $f$ itself),
that is,  \textit{Spec} can be partitioned into two sets of clauses,
call them $F_{\!\textit{\small def}}$
and $\textit{Aux\/}$, where $F_{\!\textit{\small def}}$
is the set of clauses with head predicate~$f\!,$ and~$f$ does not occur in \textit{Aux}. 

\vspace{-2mm}


\begin{theorem}[{Partial Correctness}]\label{thm:pc}
Let $F_{\!\textit{\small pcorr}}$ be the set of goals derived from 
$F_{\!\textit{\small def}}$ as follows$\,:$\ \ 
for each clause $D\!\in\! F_{\!\textit{\small def}}$ of the form 
$\textit{f\/}(X_1,\ldots,X_s,Y) \leftarrow B$,

\noindent
\hangindent=5.5mm
$(1)$~every occurrence of $f$ in $D$ 
(and, in particular, in~$B$) is replaced by $r_{\textit{\small prog}}$, 
thereby deriving a clause~$E$ of the form\/: 
$r_{\textit{\small prog}}(X_1,\ldots,X_s,Y) \leftarrow \widetilde{B}$, 

\noindent
\hangindent=5.5mm
$(2)$ clause $E$ is replaced by the goal $G$: 
$\textit{false}\leftarrow Y\!\neq\! Z,\  
r_{\textit{\small prog}}(X_1,\ldots,X_s,Z),\ \widetilde{B}$, where $Z$ is a new variable, and


\noindent
\hangindent=0mm
$(3)$  goal $G$ is replaced by the following two goals:


~~$G_1$.~ $\textit{false} \leftarrow Y\!>\! Z,\ 
r_{\textit{\small prog}}(X_1,\ldots,X_s,Z),\ \widetilde{B}$

~~$G_2$.~ $\textit{false} \leftarrow Y\!<\! Z,\ 
r_{\textit{\small prog}}(X_1,\ldots,X_s,Z),\ \widetilde{B}$


\hangindent=0mm
\noindent
Let ${\textit{PC\/}}$ be the set 
$F_{\!\textit{\small{pcorr}}} \/\cup {\textit Aux} \cup {\textit OpSem}$ of CHC's.
We have that: if ${\textit{PC}}$ is satisfiable,
then \mbox{$\{\varphi\}\,{\textit{prog}}\,\{\psi\}$} 
is valid.\eop
\end{theorem}

\vspace{-1mm}
\noindent
The proof of this theorem and of the other facts presented in this
paper can be found in the online appendix.
In our Fibonacci example  (see Example~\ref{ex:fib})
 the set  $F_{\!\textit{\small def}}$ of clauses 
is the entire set $\textit{Spec}_{\textit{\small fibonacci}}$ 
and $\textit{Aux}\!=\!\emptyset$. 
According to Points~(1)--(3) of Theorem~\ref{thm:pc},
from $\textit{Spec}_{\textit{\small fibonacci}}$ we derive the following six 
goals: 

{\small
\smallskip
\noindent
{\tt{G1. false :- F>1, r\_fibonacci(0,F).}}\nopagebreak
\hspace{10mm}

\noindent
{\tt{G2. false :- F<1, r\_fibonacci(0,F).}}\nopagebreak

\noindent
{\tt{G3. false :- F>1, r\_fibonacci(1,F).}}\nopagebreak
\hspace{10mm}

\noindent
{\tt{G4. false :- F<1, r\_fibonacci(1,F).}}

\noindent
{\tt{G5. false :- N1>=0, N2=N1+1, N3=N2+1, F3>F1+F2,}}\nopagebreak 

\hspace{18mm}{\tt{r\_fibonacci(N1,F1), r\_fibonacci(N2,F2), r\_fibonacci(N3,F3).}}
         
\noindent  
{\tt{G6. false :- N1>=0, N2=N1+1, N3=N2+1, F3<F1+F2,}}\nopagebreak

\hspace{18mm}{\tt{r\_fibonacci(N1,F1), r\_fibonacci(N2,F2), r\_fibonacci(N3,F3).}}  

} 

\noindent
Thus, in order to prove 
the validity of the specification~$(\ddagger)$ above, 
since $\textit{Aux}\!=\!\emptyset$,
it is enough to show that the set
${\textit PC\!}_{\textit{\small fibonacci}}\!=\!
\{{\texttt G1},\!\ldots,{\texttt G6}\}\/ \cup 
{\textit OpSem}_{\textit{\small fibonacci}}$ of CHC's  
is satisfiable.

\section{A Limitation of {\textit{LA}}-solving Methods}
\label{sec:Limitations}

Now we show that there are sets of CHC's that encode
partial correctness specifications whose 
satisfiability cannot be proved by {\textit {LA}}-solving methods.

A {\it symbolic interpretation} is a function $\Sigma \!: \textit{Atom} 
\longrightarrow \! \mathcal C_{{LA}}$ such that, for every $A\! \in\!
 \textit{Atom}$ and substitution $\vartheta$, 
$\Sigma(A\vartheta) = \Sigma(A)\vartheta$.
Given a set $S$ of CHC's, a symbolic interpretation $\Sigma$ is an 
${\textit{LA}}$-{\it solution}
of $S$ iff, for every clause $A_0 \leftarrow c, A_1,\ldots,A_n$ in $S$, 
we have that
${\textit{LA}} \models (c \wedge \Sigma(A_1) \wedge \ldots \wedge  
\Sigma(A_n))\rightarrow   \Sigma(A_0)$.

We say that a set $S$ of CHC's is {\textit {LA-solvable}} 
if there exists an {\textit {LA}}-solution of~$S$.
Clearly, if a  set of CHC's is {\textit {LA}}-solvable, then it is satisfiable.
The converse does not hold as we now show.

\vspace{-1mm}
\begin{theorem}\label{thm:limitation}
	There are sets of 
	constrained Horn clauses which are satisfiable and not \mbox{{\textit {LA}}-solvable}.
\end{theorem}

\vspace{-2mm}
\noindent
{\it Proof.}
Let $\textit{PC}_{{\textit{\small fibonacci}}}$ be the
set of clauses that encode the validity of the Fibonacci 
specification~$(\ddagger)$.
$\textit{PC}_{\textit{\small fibonacci}}$ is satisfiable, because
${\tt r\_fibonacci(N,F)}$ holds iff ${\tt F}$ is the
${\tt N}$-th Fibonacci number, and hence the bodies of 
${\tt G1}, \ldots, {\tt G6}$ are false.
(This fact will also be proved by the automatic method presented in 
Section~\ref{sec:transform}.)

Now we prove, by contradiction, that 
$\textit{PC}_{\textit{\small fibonacci}}$ is not {\textit {LA}}-solvable.
Suppose that there exists an {\textit {LA}}-solution $\Sigma$ 
of $\textit{PC}_{{\textit{\small fibonacci}}}$.
Let $\Sigma ({\tt r\_fibonacci(N,F)})$ be a constraint ${\tt c(N,F)}$.
To keep our proof simple, we assume that ${\tt c(N,F)}$ is defined by a 
conjunction of linear arithmetic inequalities (that is, ${\tt c(N,F)}$ 
is a convex constraint), 
but our argument can easily be generalized to any constraint in~$\mathcal C_{{LA}}$.
By the definition of {\textit {LA}}-solution, we have that:


{\small{
\noindent
$(P1)$ {{${\textit{LA}} \not\models \exists({\tt N1\!>=\!0, 
N2\!=\!N1\!+\!1, N3\!=\!N2\!+\!1, F3\!>\!F1\!+\!F2, c(N1,F1), 
c(N2,F2), c(N3,F3)})$}} 
		
\noindent
$(P2)$ {{$M(\textit{OpSem}_{\textit{\small fibonacci}}) \models 
\forall\ ( {\tt r\_fibonacci(N,F) \rightarrow  c(N,F)})$}} }}
	
\noindent
Property~$(P1)$ follows from the fact that, in particular, an 
\textit{LA}-solution satisfies goal~{\tt G5}.
Property~$(P2)$ follows from the fact that an \textit{LA}-solution 
satisfies all clauses of 
$\textit{OpSem}_{\textit{\small fibonacci}}$ and
$M(\textit{OpSem}_{\textit{\small fibonacci}})$ defines the 
{\it least\/} ${\tt r\_fibonacci}$ relation that satisfies 
those clauses.

\begin{figure}[!ht]
\vspace{-2mm}
\centering
\includegraphics[50mm,209mm][160mm,252mm]{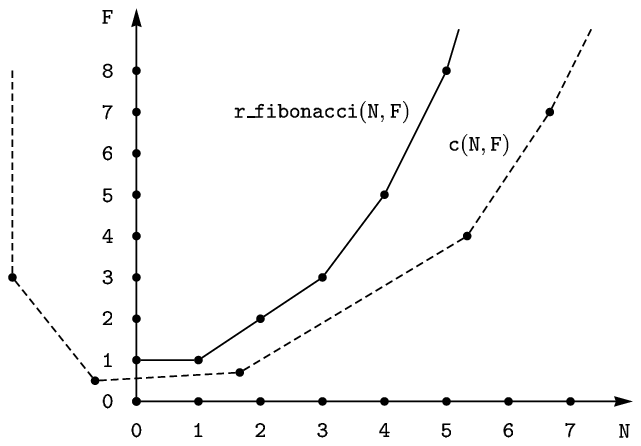}
\caption{The relation {\tt r\_fibonacci(N,F)} and the convex constraint 
{\tt c(N,F)}.\label{fig:fib-constraint}
}
\vspace{-2mm}
\end{figure}

From Property $(P2)$ and from the fact that ${\tt r\_fibonacci(N,F)}$ 
holds iff ${\tt F}$ is the
${\tt N}$-th Fibonacci number (and hence {\tt F} is an exponential 
function of {\tt N}), 
it follows that {\tt c(N,F)} is a conjunction of the form 
${\tt c_1(N,F), \ldots,  c_k(N,F)}$,
where, for ${\tt i=1,\ldots,k}$, with ${\tt k\!\geq\!0}$, ${\tt c_i(N,F)}$ is
either 
(A)~${\tt N\! >\! a_i}$, for some integer ${\tt a_i}$, or 
(B)~${\tt F\! >\! a_i\! \cdot\! N\! +\! b_i}$. 
(No constraints of the form ${\tt F\! <\! a_i\! \cdot\! N\! +\! b_i}$
are possible, as shown in Figure~\ref{fig:fib-constraint}.)

\noindent
By replacing {\tt c(N1,F1)}, {\tt c(N2,F2)}, and {\tt c(N3,F3)}
by the corresponding conjunctions of atomic constraints of the forms (A) and (B), and eliminating the occurrences of
{\tt F1}, {\tt F2}, {\tt N2}, and {\tt N3}, from $(P1)$ we get:

{\small{
\noindent
$(P3)$ ${\textit{LA}}\! \not\models\! \exists ( {\tt N1{\geq}0, F3\!>\!p_1, \ldots, F3\!>\!p_n})$
}}

\noindent
where, for ${\tt i=1,\ldots,n}$, ${\tt p_i}$ is a linear polynomial in the variable  {\tt N1}.
Then, the constraint ${\tt N1{\geq}0, F3\!>\!p_1, \ldots, F3\!>\!p_n}$ is satisfiable
and Property~$(P3)$ is false. Thus, 
the assumption 
that  $\textit{PC}_{\textit{\small fibonacci}}$ is {\textit {LA}}-solvable is false,
and we get the thesis.
\hfill\eop

\section{Increasing the Power of LA-Solving Methods by Linearization}
\label{sec:transform}

A weakness of the {\it LA}-solving methods
is that they look for $\textit{LA}$-solutions constructed from single atoms, 
and by doing so they may fail to discover that a goal is satisfiable 
because a conjunction of atoms in its premise is unsatisfiable,
in spite of the fact that {each of its conjoint atoms is} satisfiable.
For instance, in our Fibonacci example  the premise of 
goal~{\tt G5} contains three atoms with predicate {\tts r\_fibonacci} and our proof
of Section \ref{sec:Limitations} shows that, even if the premise of~{\tt G5}
is unsatisfiable, there is no constraint which is an $\textit{LA}$-solution of the 
clauses defining
{\tts r\_fibonacci} that, when substituted for each {\tts r\_fibonacci} atom, 
makes that premise false.
Thus, the notion of $\textit{LA}$-solution
shows some weakness when dealing with 
{\it nonlinear} clauses, that is, clauses whose premise 
contains more than one atom (besides constraints).

In this section we present an automatic transformation of
constrained Horn clauses that has the objective of increasing the power of
 ${\textit{LA}}$-solving methods. 
 
The core of the transformation, called {\it linearization}, 
takes a set of possibly 
nonlinear constrained Horn clauses and transforms it
 into a set of {\it linear} clauses,
that is, clauses whose premise contains at most one atom (besides constraints).
After performing linearization, the {\it LA}-solving methods are able
to exploit the interactions among several atoms,
instead of dealing with each atom individually.
In particular, an ${\textit{LA}}$-solution of the linearized set of clauses
will map a {\it conjunction} of atoms
to a constraint.
We will show that {\rm linearization} preserves the existence of 
 ${\textit{LA}}$-solutions and,
in some cases (including our Fibonacci example), transforms a set of clauses 
which is not ${\textit{LA}}$-solvable into a set of clauses that is 
${\textit{LA}}$-solvable.

Our transformation technique is made out of the following two steps:\\
(1)~RI:~{\it Removal of the interpreter}, and
(2)~LIN:~{\it Linearization}.\\ 
These steps
 are performed by using the 
transformation rules for CLP programs presented in \cite{EtG96}, that is:
 {\it unfolding} (which consists in applying a resolution step
and a constraint satisfiability test), {\it definition} (which 
introduces a new predicate defined in terms of old predicates), 
and {\it folding} (which redefines old predicates
in terms of new predicates introduced by the definition rule).

\subsection{{\bf{RI:}} Removal of the Interpreter}
\label{subsec:removal}

This step is a variant of the  {removal of the interpreter}
transformation presented in~\cite{De&14c}.
In this step a specialized definition for  $r_{\textit{prog}}$ is derived
by transforming the CLP program \textit{OpSem}, thereby getting a new
CLP program $\textit{OpSem}_{\textit{\small{RI}}}$ where there are
no occurrences of the predicates {\it initCf}, {\it finalCf},
{\it reach}, and~{\it tr}, which as already mentioned 
encodes the interpreter of the imperative language in 
which {\it prog} is written.
(See online appendix for more details.)

By a simple extension of the results presented in~\cite{De&14c},
it can be shown that the RI transformation always terminates, 
preserves satisfiability, and transforms \textit{OpSem} into a set of
linear clauses over the integers.
It can also be shown that the removal of the interpreter preserves 
\mbox{${\textit{LA}}$-solvability.} 
	Thus, we have the following result.

\vspace{-.5mm}
\begin{theorem}
Let \textit{OpSem} be a CLP program constructed starting from any given 
imperative program \textit{prog}.\label{thm:RI}
Then the RI transformation terminates and derives a CLP program $\textit{OpSem}_{\textit{\small{RI}}}$ such that:

\noindent
(1) $\textit{OpSem}_{\textit{\small{RI}}}$ is a set of linear clauses over the integers;

\noindent
(2) $\textit{OpSem} \cup \textit{Aux}\, \cup F_{\textit{\small{pcorr}}}$ is satisfiable iff 
$\textit{OpSem}_{\textit{\small{RI}}} \cup \textit{Aux}\, \cup F_{\textit{\small{pcorr}}}$ is satisfiable;

\noindent
(3) $\textit{OpSem}\, \cup \textit{Aux}\, \cup F_{\textit{\small{pcorr}}}$ is 
${\textit{LA}}$-solvable iff $\textit{OpSem}_{\textit{\small{RI}}}\, \cup 
\textit{Aux}\, \cup F_{\textit{\small{pcorr}}}$ is {\textit{LA}}-solvable.
\end{theorem}

\noindent
In the Fibonacci example, the input of the
RI transformation is $\textit{OpSem}_{\textit{fibonacci}}$. 
The output of the
RI transformation consists of the following three clauses:

{\small{
\noindent
{\tt{E1.~r\_fibonacci(N,F):- N>=0,\,U=1,\,V=0,\,T=0,\,r(N,U,V,T,N1,F,V1,T1).}}\rule{0mm}{3.5mm}

\noindent
{\tt{E2.~r(N,U,V,T,N,U,V,T):- N=<0.}}

\noindent
{\tt{E3.~r(N,U,V,T,N2,U2,V2,T2):- N>=1,\,N1=N-1,\,U1=U+V,\,V1=U,\,T1=U,}}

\hspace{46mm}{\tt{r(N1,U1,V1,T1,N2,U2,V2,T2).}}

}} 
\noindent
where {\tts r} is a new predicate symbol introduced by the
RI transformation.

As stated by Theorem~\ref{thm:RI}, $\textit{OpSem}_{\textit{\small{RI}}}$ is a set of
clauses over the integers. Since the clauses of the specification {\it Spec} define
computable functions from $\mathbb Z ^s$ to $\mathbb Z$, without loss of generality we may assume that also
the clauses in $\textit{Aux}\, \cup F_{\textit{\small{pcorr}}}$ are over the integers 
\cite{SeS82}.
From now on we will only deal with clauses over the integers, and
we will feel free to omit the qualification `over the integers'.

\subsection{{\bf{LIN:}} Linearization}
\label{subsec:linearization}

The {\rm linearization} transformation takes as input the set 
$\textit{OpSem}_{\textit{\small{RI}}} \cup \textit{Aux}\, \cup F_{\textit{\small{pcorr}}}$
of constrained Horn clauses and derives a new, equisatisfiable set \textit{TransfCls} of 
{\it linear} constrained Horn clauses. 

In order to perform linearization, we assume that
\textit{Aux} is a set of linear clauses.
This assumption,
which is not restrictive because any computable function on the integers can be
encoded by linear clauses \cite{SeS82}, simplifies the 
proof of termination of the transformation.

The {\rm linearization} transformation is described in Figure~\ref{fig:Lin}.
Its input is constructed  by partitioning
$\textit{OpSem}_{\textit{\small{RI}}} \cup \textit{Aux}\, \cup F_{\textit{\small{pcorr}}}$
into a set {\it LCls} of linear clauses and a set {\it NLGls} of nonlinear goals.
{\it LCls} consists of \textit{Aux}, $\textit{OpSem}_{\textit{\small{RI}}}$ 
(which, by Theorem~\ref{thm:RI}, is a set of linear clauses),
and the subset of linear goals in $F_{\textit{\small{pcorr}}}$. 
{\it NLGls} consists of the set of nonlinear goals in $F_{\textit{\small{pcorr}}}$.

When applying linearization we use the following transformation rule.

\smallskip
\noindent
{\it Unfolding  Rule.} Let $\textit{Cls}$ be a set of constrained Horn clauses.
Given a clause $C$ 
of the form $H\leftarrow c,\textit{Ls},A,\textit{Rs}$, 
let us consider the set
$\{{K}_i \leftarrow {c}_i,B_i \mid  i=1, \ldots, m\}$ 
made out of the (renamed apart) clauses of~$\textit{Cls}$ 
such that, for $i\!=\!1,\ldots,m,$ 
${A}$ is unifiable with~${K}_i$ via the most general 
unifier~$\vartheta_i$ and $({c,c}_i)\, \vartheta_i$ is satisfiable. 
By unfolding~$C$ with respect to~$A$ using~$\textit{Cls}$, we derive the set 
$\{({H}\!\leftarrow c,c_i,\textit{Ls},B_i,\textit{Rs})\,\vartheta_i\! \mid
 i\!=\!1, \ldots, m\}$ of clauses.

\begin{figure}[ht]
\vspace{-3mm}

\noindent\hrulefill
\begin{flushleft}\vspace{-2mm}
\noindent {\it Input\/}: (i)~A set \textit{LCls} of linear clauses, 
and (ii)~a set \textit{Gls} of nonlinear goals.\\ 
\noindent  {\it Output\/}: A set \textit{TransfCls} of linear clauses.

\vspace*{-2mm}
\rule{30mm}{0.1mm}

\noindent \textsc{Initialization}:~~~
$\textit{NLCls}:= \textit{Gls}$;
~ ~$\textit{Defs}:= \emptyset$;
~ ~$\textit{TransfCls}:= \textit{LCls}$;

\smallskip

\noindent \textit{while}~ there is a clause~$C$ in \textit{NLCls}
~\textit{do}

\smallskip
\hspace*{3mm}\begin{minipage}{124mm} 

\hangindent=3mm
\noindent \textsc{Unfolding}: From clause $C$ derive a set $\textit{U}(C)$ of clauses by
unfolding $C$ with respect to every atom occurring
in its body using \textit{LCls}; \\
Rewrite each clause in $\textit{U}(C)$ to a clause of the form 
 $H \leftarrow c,\, A_1, \ldots, A_k$, such that, for $i=1,\ldots,k$, $A_i$ is of the form
$p(X_1,\ldots,X_m)$;

\smallskip
\hangindent=3mm
\noindent {\textsc{Definition}\,\&\,\textsc{Folding}:} 

\noindent
\hspace*{3mm}$F(C) := U(C)$;

\hangindent=3mm
\noindent
\hspace*{3mm}\makebox[4mm][l]{\it for} every clause $E\in F(C)$ of the form  $H \leftarrow c,\, A_1, \ldots, A_k$ \ {\it do}

\hangindent=11.5mm
\noindent
\hspace*{8mm}{\it if} in \textit{Defs} there is no
clause of the form  $\textit{newp}(X_1,\ldots,X_t) \leftarrow A_1, \ldots, A_k$,
where $\{X_1,\ldots,X_t\} = \textit{vars}(A_1, \ldots, A_k) \cap \textit{vars}(H,c)$

\noindent
\hangindent=18mm
\hspace*{8mm}{\it then} add $\textit{newp}(X_1,\ldots,X_t) \leftarrow A_1, \ldots, A_k$ to \textit{Defs} and to \textit{NLCls}; 

\noindent
\hspace*{8mm}$F(C) := (F(C) - \{E\}) \cup \{H \leftarrow c,\, \textit{newp}(X_1,\ldots,X_t)\}$\\[-5mm]

\noindent
\hspace*{3mm}\makebox[5mm][l]{\it end-for}

\smallskip
\noindent $\textit{NLCls}:=\textit{NLCls}-\{C\}$;
~~~ $\textit{TransfCls}:=\textit{TransfCls}\cup \textit{F}(C)$;

\end{minipage} 

\smallskip

\noindent \textit{end-while}

\end{flushleft}

\vspace*{-4mm}
\noindent\hrulefill
\vspace*{-2.mm}
\caption {LIN: The \textrm{linearization} transformation.}
\label{fig:Lin}
\vspace*{-4mm}

\end{figure}

It is easy to see that, since \textit{LCls} is a set of  linear clauses, 
only a finite number of new predicates can be 
 introduced by any sequence of applications of
\textsc{Definition}\,\&\,\textsc{Folding}, and hence 
the {\rm linearization} transformation terminates.
Moreover, the use of the unfolding, definition, and folding rules 
according to the conditions indicated in \cite{EtG96},
guarantees the equivalence with respect to the least ${\textit{LA}}$-model, and hence
the equisatisfiability of $\textit{LCls}\, \cup \textit{Gls}$ and \textit{TransfCls}. 
Thus, we have the following result. 
\vspace{-1mm}

\begin{theorem}[{\textit Termination and Correctness of Linearization}]\label{thm:term-corr}
Let {\it LCls} be a set of linear clauses and \textit{Gls} be a set of nonlinear goals. The \textrm{linearization} transformation terminates for the input set of clauses $\textit{LCls}\, \cup \textit{Gls}$, and the output $\textit{TransfCls}$ is a set of linear clauses.
Moreover, $\textit{LCls}\, \cup \textit{Gls}$ is satisfiable iff
 \textit{TransfCls} is satisfiable.~\eop
\end{theorem}

\noindent
Let us consider again the Fibonacci example.
We apply the {\rm linearization} transformation to the set $\{${\tts E1,E2,E3}$\}$
of linear clauses, and to the nonlinear goal {\tts G5}. For brevity, we omit to
consider the cases where the goals {\tt G1$,\ldots,$G4$,$G6} are taken as input to the {\rm linearization} transformation.

After \textsc{Initialization} we have that
$\textit{NLCls}=\{${\tts G5}$\}$, $\textit{Defs}=\emptyset$, and 
$\textit{TransfCls}=\{${\tts E1,E2,E3}$\}$.
By applying the \textsc{Unfolding} step to {\tts G5} we derive:

{\small{
\noindent
\makebox[7mm][l]{{\tt C1}.}\makebox[15mm][l]{{\tt false :-}}{\tt N1>= 0, N2=N1+1, N3=N2+1, F3>F1+F2, U=1, V=0,}\nopagebreak

\noindent
\hspace{22mm}{\tt{r(N1,U,V,V,X1,F1,Y1,Z1), r(N2,U,V,V,X2,F2,Y2,Z2), }}\nopagebreak

\noindent
\hspace{22mm}{\tt{r(N3,U,V,V,X3,F3,Y3,Z3).}}

}} 

\noindent
Next, by \textsc{Definition}\,\&\,\textsc{Folding}, the following clause is added to 
\textit{NLCls} and \textit{Defs}:

{\small{ 
\noindent
\makebox[7mm][l]{{\tt C2}.}\makebox[10mm][l]{{\tt new1(N1,U,V,F1,N2,F2,N3,F3) :- r(N1,U,V,V,X1,F1,Y1,Z1), }}

\noindent
\hspace{22mm}{\tt{r(N2,U,V,V,X2,F2,Y2,Z2), r(N3,U,V,V,X3,F3,Y3,Z3).}}

}} 

\noindent
and clause {\tt C1} is folded using  {\tt C2}, thereby deriving the following linear clause:

{\small{ 
\noindent
\makebox[7mm][l]{{\tt C3}.}\makebox[15mm][l]{{\tt false :-}}{\tt N1>= 0, N2=N1+1, N3=N2+1, F3>F1+F2, U=1, V=0,}

\noindent
\hspace{22mm}{\tt{new1(N3,U,V,F3,N2,F2,N1,F1).}}

}} 

\noindent
At the end of the first execution of the body of the \textit{while-do} loop we have: 
$\textit{NLCls}=\{${\tts C2}$\}$, $\textit{Defs}=\{${\tts C2}$\}$, and 
$\textit{TransfCls}=\{${\tts E1,E2,E3,C3}$\}$. Now, the {\rm linearization} 
transformation continues by processing clause {\tts C2}.
During its execution, {\rm linearization} introduces two new predicates defined by
the following two clauses:

{\small{ 
\noindent
\makebox[7mm][l]{{\tt C4}.}{\tt new2(N,U,V,F)\,:-\,r(N,U,V,V,X,F,Y,Z).}

\noindent
\makebox[7mm][l]{{\tt C5}.}{\tt new3(N2,U,V,F2,N1,F1)\,:-\,r(N1,U,V,V,X1,F1,Y1,Z1),\,r(N2,U,V,V,X2,F2,Y2,Z2).}

}} 

The transformation terminates 
when all clauses derived by unfolding can be folded using clauses in
\textit{Defs}, without introducing new predicates.
The output of the transformation is a set of linear clauses (listed in the
online appendix) which is {\it LA}-solvable, as reported on line 4 of Table \ref{tab:expres} in the 
next section.

In general, there is no guarantee that we can automatically transform any given satisfiable set of clauses into an {\it LA}-solvable one. In fact, such a transformation cannot be algorithmic because, for constrained Horn clauses, the problem of satisfiability is not semidecidable, while the problem of {\it LA}-solvability is semidecidable (indeed, the set of symbolic interpretations is recursively enumerable and the problem of checking whether or not a symbolic interpretation is an {\it LA}-solution is decidable).
However, the {\rm linearization} transformation cannot decrease {\it LA}-solvability, as the following theorem shows. 

\vspace{-2mm}
\begin{theorem}[Monotonicity with respect to~{\it LA}-Solvability]
\label{thm:LAsolv}Assume that by applying
the {\rm linearization} transformation to a set $\textit{LCls}\cup \textit{Gls}$
of CHC's, we obtain a set $\textit{TransfCls}$. 
If $\textit{LCls}\cup \textit{Gls}$ is
{\it LA}-solvable, then $\textit{TransfCls}$ is {\it LA}-solvable.\eop
\end{theorem}

\vspace{-2mm}

Since there are cases where
$\textit{LCls}\cup \textit{Gls}$ is not \mbox{{\it LA}-solvable},
while $\textit{TransfCls}$ is  \mbox{{\it LA}-solvable} 
(see the Fibonacci example above
and some more examples in the following section), as a consequence of 
Theorem \ref{thm:LAsolv} we get that the combination of {\it LA}-solving and
{\rm linearization} is strictly 
more powerful than {\it LA}-solving alone.

\section{Experimental Results}
\label{sec:experiments}
\begin{figure}[!ht]
\vspace{-2mm}
\centering
\scalebox{1.47}{\includegraphics[2mm,3.5mm][90mm,17mm]{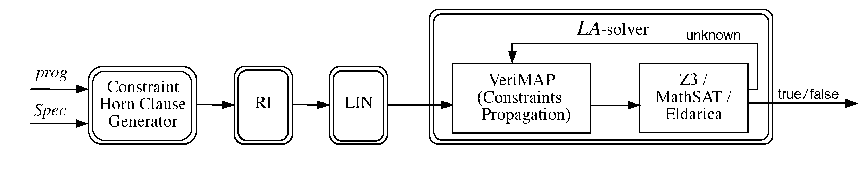}}
\vspace{-4mm}
\caption{Our software model checker that uses the linearization module LIN.\label{fig:RI-LIN-LASolver}
}
\vspace{-3mm}
\end{figure}

We have implemented our verification method by using the VeriMAP system~\cite{De&14b}.
The implemented tool consists of four modules, which we have depicted in 
Figure~\ref{fig:RI-LIN-LASolver}. The first module, given the imperative 
program {\it prog} and its specification {\it Spec}, generates the set {\it PC}
of constrained Horn 
clauses  (see Theorem~\ref{thm:pc}). {\it PC\/} is then 
given as input to the module RI  that removes the 
interpreter. Then, the module LIN  
performs the {\rm linearization}  transformation. 
Finally, the resulting linear clauses are passed to the {\it LA}-solver, consisting
of \mbox{VeriMAP} together with an SMT solver, which is either 
Z3~\cite{DeB08} 
or MathSAT~\cite{MaS13} or Eldarica~\cite{Ru&13}.

\vspace*{-2mm}

We performed an experimental evaluation
on a set of programs taken from the literature,
including some programs from~\cite{Fe&14}
obtained by applying 
strength reduction, 
a real-world optimization technique\footnote{{\tt https://www.facebook.com/notes/facebook-engineering/three-optimization-tips-for-c/\\10151361643253920}}.
In Table~\ref{tab:expres}
we report the results of our experiments\footnote{The VeriMAP tool, source code and specifications for the programs are available at: 
{\tt http://map.uniroma2.it/linearization}}. 

One can see that {\rm linearization} takes very little time
compared to the total verification time.
Moreover, {\rm linearization} is necessary for the verification 
of 14 out of 19 programs (including {\it fibonacci\/}), which otherwise 
cannot be proved correct with respect to their specifications.
In the two columns under {\it LA}-solving-1 
we report the results obtained by giving as input to the  Z3 and 
Eldarica solvers the set~\textit{PC} generated by the RI module.
Under {\it LA}-solving-1 we do not have a column for MathSAT, 
because the version of this solver used in our experiments 
(namely, MSATIC3) cannot deal with nonlinear CHC's, 
and therefore it cannot be applied before linearization.
In the last three columns of Table~\ref{tab:expres} we report the
results obtained by giving as input to VeriMAP (and the solvers Z3, 
MatSAT, and Eldarica, respectively) the clauses obtained after linearization.

Unsurprisingly, for the verification problems where linearization is not necessary, 
our technique may deteriorate the performance, although in most of these problems
the solving time does not increase much.


\begin{table}[ht]
\vspace{-2mm}
\begin{tabular}
{|@{\hspace{1mm}}r@{\hspace{-1mm}}l@{\hspace{6pt}}||
	@{\hspace{-1pt}}r@{\hspace{4pt}}| 
	@{\hspace{-6pt}}r|@{\hspace{-4pt}}r@{\hspace{3pt}}
	@{\hspace{0pt}}||@{\hspace{0pt}}r@{\hspace{4pt}}| 
	@{\hspace{-6pt}}r@{\hspace{2pt}}|@{\hspace{-4pt}}r|@{\hspace{-4pt}}r@{\hspace{3pt}}|}

\cline{1-9} 
& 
\multirow{2}{*}{{\it Program} } & 
\multirow{2}{*}{{\rm RI~}} &
\multicolumn{2}{c||@{\hspace{0pt}}}{{\rule{0mm}{3.8mm}}{\it{LA}}{\rm{-solving-1}}} &
\multirow{2}{*}{\rm LIN}  & 
\multicolumn{3}{c|}{{\rule{0mm}{3.5mm}}{\it{LA}}{\rm{-solving-2: VeriMAP \&}}}\\ 
\cline{7-9}  
\cline{4-5}
& & & {\rm Z3~} & {\rm Eldarica} &
&\rule{0mm}{3.5mm}{\rm Z3~~} 
&{\rm MathSAT} 
&{\rm Eldarica} \\[0pt]
\cline{1-9} \\[-9.5pt]\cline{1-9}

1.& {\it binary\_division\rule{0mm}{3.mm}} & 0.02 
& ~4.16 & {\it TO} 
& 0.04 
& ~17.36 & 17.87 & 20.98 \\[0pt]

2.& {\it fast\_multiplication\_$2$\rule{0mm}{0mm}} & 0.02 
& {\it TO} & 3.71 
& 0.01 
& 1.07 & 1.97 & 7.59\\[0pt]

3.& {\it fast\_multiplication\_$3$\rule{0mm}{0mm}} & 0.03 
& {\it TO} & 4.56 
& 0.02 
& 2.59 & 2.54 & 9.31 \\[0pt]

4.& {\it fibonacci \rule{0mm}{0mm}} & 0.01 
& {\it TO} & {\it TO} 
& 0.01 
& 2.00 & 47.74 & 6.97\\[0pt]

5.& {\it Dijkstra\_fusc\rule{0mm}{0mm}} & 0.01 
& 1.02 & 3.80 
& 0.05 
& 2.14 & 2.80 & 10.26 \\[0pt]

6.& {\it greatest\_common\_divisor\rule{0mm}{0mm}} & 0.01 
& {\it TO} & {\it TO} 
& 0.01 
& 0.89 & 1.78 & 0.04\\[0pt]

7.& {\it integer\_division\rule{0mm}{0mm}} & 0.01 
& {\it TO} & {\it TO} 
& 0.01 
& 0.88 & 1.90 & 2.86\\[0pt]

8.  &{\it $91$-function\rule{0mm}{0mm}} & 0.01 
& 1.27  & {\it TO} 
& 0.06 
& ~117.97 & 14.24 & {\it TO} \\[0pt]

9.& {\it integer\_multiplication\rule{0mm}{0mm}} & 0.02
& {\it TO}  & {\it TO} 
& 0.01
& 0.52 & 14.76 & 0.54\\[0pt]

10. & {\it remainder\rule{0mm}{0mm}} & 0.01 
& {\it TO}  & {\it TO} 
& 0.01 
& 0.87 & 1.70 & 3.16\\[0pt]

11. & {\it sum\_first\_integers\rule{0mm}{0mm}} & 0.01 
& {\it TO} & {\it TO} 
& 0.01 
& 1.79 & 2.30 & 6.81\\[0pt] 

12. & {\it lucas\rule{0mm}{0mm}} & 0.01 
& {\it TO} & {\it TO} 
& 0.01 
& 2.04  & 8.39 & 9.46\\[0pt]  

13. & {\it padovan\rule{0mm}{0mm}} & 0.01 
& {\it TO} & {\it TO} 
& 0.01
& 2.24  & {\it TO} & 11.62\\[0pt] 

14. & {\it perrin\rule{0mm}{0mm}} & 0.01 
& {\it TO} & {\it TO} 
& 0.02 
& 2.23 & {\it TO} & 11.89\\[0pt] 

15. & {\it hanoi\rule{0mm}{0mm}} & 0.01 
& {\it TO} & {\it TO} 
& 0.01 
& 1.81 & 2.07 & 6.59\\[0pt]

16. & {\it digits$10$ \rule{0mm}{0mm}} & 0.01
& {\it TO} & {\it TO} 
& 0.01 
& 4.52 & 3.10 & 6.54 \\[0pt]

17. & {\it digits$10$-itmd \rule{0mm}{0mm}} & 0.06 
& {\it TO} & {\it TO} 
& 0.04 
& {\it TO} & 10.26 & 12.38 \\[0pt]

18. & {\it digits$10$-opt \rule{0mm}{0mm}} & 0.08
& {\it TO} & {\it TO} 
& 0.10 
& {\it TO} & {\it TO} & 15.80\\[0pt]

19. & {\it digits$10$-opt\/$100$ \rule{0mm}{0mm}} & 0.01
& {\it TO} & {\it TO} 
& 0.02 
& {\it TO} & 58.99 & 8.98 \\[0pt]

\cline{1-9} 
\end{tabular}\vspace{0mm}

\caption{Columns\/ {\rm RI} and {\rm LIN} show the times $($in seconds$)$
 taken for removal of the interpreter and linearization. 
The two columns under \mbox{{\it LA}-{\rm solving-1}} show the times taken by {\rm Z3} and {\rm \hspace{.5mm}Eldarica} for solving the problems after {\rm RI} alone.
The three columns under \mbox{{\it LA}-{\rm solving-2}} show the times taken 
by \mbox{\rm VeriMAP} together with {\rm Z3}, {\rm MathSAT}, {\rm \hspace{.5mm}and Eldarica},  after {\rm RI} and {\rm LIN}.
The timeout TO occurs after $120$ seconds.
\label{tab:expres}}
\end{table}


\section{Conclusions and Related Work}
\label{sec:conclusions}

We have presented a method for proving partial correctness specifications
of programs, given as Hoare triples of the form $\{\varphi\}\, \textit{prog}\, \{\psi\}$,
where the assertions~$\varphi $ and~$\psi$ are predicates
defined by a set of {\it possibly recursive}, definite CLP clauses.
Our verification method is based on: 
{\it{ Step}} (1) a translation of a given
specification into a set of constrained 
Horn clauses (that is, a CLP program together with one or more goals), 
{\it{Step}} (2)  an unfold/fold transformation strategy, 
called {\rm linearization}, which
derives {\it linear} clauses (that is, clauses with at most 
one atom in their body), and
{\it{Step}} (3) an {\it LA}-solver that attempts to prove the 
satisfiability of constrained Horn clauses
by interpreting predicates as linear arithmetic constraints.

We have formally proved that the method {which uses} {\rm linearization} 
is strictly more
powerful than the method that applies Step\,(3) immediately after Step\,(1).
We have also developed a proof-of-concept implementation of our method 
by using the VeriMAP verification system~\cite{De&14b} together with 
various state-of-the-art solvers
(namely, Z3~\cite{DeB08}, MathSAT~\cite{MaS13}, and Eldarica~\cite{Ru&13}), and we 
have shown that our method works on several verification problems.
Although these problems refer to
quite simple specifications, some of them cannot be solved by {using the above mentioned solvers alone.}

The use of transformation-based methods in the field of program verification 
has recently gained  popularity  (see, 
for instance,~\cite{Al&07,De&14c,Fi&13a,KaG15a,LeM99,LiN08,Pe&98}). 
However, fully automated methods based on various notions
of {\it partial deduction} and {\it CLP program specialization} cannot
achieve the same effect as {\rm linearization}. Indeed, 
{\rm linearization} requires the introduction of new predicates corresponding
to {\it conjunctions} of old {predicates}, whereas {partial deduction} and {program specialization}
can only introduce new predicates that correspond to instances of old 
{predicates}.
In order to derive linear clauses, one could apply {\it conjunctive partial deduction}
\cite{De&99}, which essentially is equivalent to unfold/fold transformation. However, 
to the best of our knowledge, this 
application of conjunctive partial deduction to the field of program verification
has not been investigated {so far}.

The use of linear arithmetic constraints for program verification
has been first proposed in the field of 
{\it abstract interpretation}~\cite{CoC77}, where
these constraints are used for approximating the set of states that
are reachable during program execution~\cite{CoH78}.
In the field of logic programming, abstract interpretation methods work similarly to 
{\it LA}-solving for constrained Horn clauses, 
because they both 
look for interpretations of predicates as linear arithmetic 
constraints that satisfy {the} program clauses 
(see, for instance,~\cite{BeK96}).
Thus, abstract interpretation methods suffer from the 
same theoretical limitations we 
have pointed out in this paper for {\it LA}-solving methods.

One approach that has been followed for overcoming the limitations
related to the use of linear arithmetic constraints is to
devise methods for generating polynomial invariants and proving
specifications with polynomial arithmetic constraints~\cite{RoK07a,RoK07b}. 
This approach also requires the development
of solvers for polynomial constraints, which is a very complex task on its own,
as in general the satisfiability of these constraints on the integers 
is undecidable~\cite{Mat70}.
In contrast, the approach presented in this paper  has the objective of transforming
problems which would require the proof of nonlinear 
arithmetic assertions into problems
which can be solved by using linear arithmetic constraints.
We have shown some examples (such as the  {\it fibonacci} program)
where we are able to prove specifications whose post-condition
is an exponential function.

An interesting issue for future research is 
to identify general criteria to answer the following question:
Given a class $\mathcal D$ of constraints and a class $\mathcal H$
of constrained Horn clauses, does the satisfiability
of a finite set of clauses in $\mathcal H$ 
imply its \mbox{$\mathcal D$-solvability?}
Theorem \ref{thm:limitation} provides a negative answer to this question when
$\mathcal D$ is the class of {\it LA} constraints and
$\mathcal H$ is the class of all constrained Horn clauses.


\section{Acknowledgments}

We thank the participants in the Workshop VPT~'15 on
{\it Verification and Program Transformation}, held in London on April 2015, 
for their comments on a preliminary version of this paper.
This work has been partially supported by the National Group of 
Computing Science (GNCS-INDAM).


\newpage

\section*{Appendix}

For the proof of Theorem \ref{thm:pc} we need the following lemma.

\medskip
\noindent
{\it Lemma~1}. (i) The relation $r_{\textit{\small prog}}$ defined by \textit{OpSem} is a functional relation, that is, 
$M(\textit{OpSem}) \models \forall p_1,\!\ldots\!,p_s,y_1,y_2$.$r_{\textit{\small prog}}(p_1,\!\ldots\!,p_s,y_1) 
\!\wedge\! r_{\textit{\small prog}}(p_1,\!\ldots\!,p_s,y_2)\! \rightarrow\! y_1\!=\!y_2$.

\noindent
(ii) A program
${\textit{prog}}$ terminates for an environment $\delta_0$ such that
\mbox{$\delta_{0}(z_{1})\!=\!p_{1}$},$\ldots$, $\delta_{0}(z_{s})\!=\!p_{s}$
and $\textit{pre}(p_1,\!\ldots\!,p_s)$ holds, iff

$M(\textit{OpSem}) \models \textit{pre}(p_1,\!\ldots\!,p_s) 
 \rightarrow \exists y$. $r_{\textit{\small prog}}(p_1,\!\ldots\!,p_s,y) $.

\medskip

\noindent
{\it Proof.} 
Since the program \textit{prog}  is deterministic, 
the predicate $r_{\textit{\small prog}}$ defined by \textit{OpSem} is 
a functional relation (which might not be total on \textit{pre}, as 
\textit{prog} might not terminate).
Moreover, a program \textit{prog}, with variables $z_{1},\ldots,z_{s}$,
terminates for an environment $\delta_{0}$ such that:
(i)~$\delta_{0}(z_{1})\!=\!p_{1},$ $\ldots,$ $\delta_{0}(z_{s})\!=\!p_{s}$,
and (ii)~$\delta_{0}$  satisfies~\textit{pre},
iff  $\exists y$.\,$r_{\textit{\small prog}}(p_1,\ldots,p_s,y) $ holds in 
$M(\textit{OpSem})$. \hfill\eop

\medskip

\noindent
{\it Proof of Theorem \ref{thm:pc} $($Partial Correctness$)$.}

\noindent
Let $\textit{dom}_r(X_1,\ldots,X_s)$ be  a predicate that
represents the {\it domain} of the functional relation $r_{\textit{prog}}$.
We assume that   $\textit{dom}_r(X_1,\ldots,X_s)$ is defined by
a set \textit{Dom} of clauses, using predicate symbols not in $\textit{OpSem} \cup \textit{Spec}$, such that 

$M(\textit{OpSem}\cup \textit{Dom})\models$ \hfill $(1)$

\hspace{20mm}$\forall X_1,\ldots,X_s\textit{.} ( (\exists Y\textit{.}  r_{\textit{prog}}(X_1,\ldots,X_s,Y) \leftrightarrow \textit{dom}_r(X_1,\ldots,X_s))$ 

\noindent
Let us denote by $\textit{Spec}^{\sharp}$ the set of clauses obtained from 
\textit{Spec} by replacing each clause  $\textit{f\/}(X_1,\ldots,X_s,Y) \leftarrow B$
by the clause $\textit{f\/}(X_1,\ldots,X_s,Y) \leftarrow \textit{dom}_r(X_1,\ldots,X_s), B$.
Then,  for all integers $p_1,\ldots,p_s,y$,

\smallskip

$M(\textit{Spec}^{\sharp} \cup \textit{Dom}) \models \textit{f\/}(p_1,\ldots,p_s,y) $ \ implies \ $M(\textit{Spec}) \models \textit{f\/}(p_1,\ldots,p_s,y) $  \hfill (2)

\smallskip

\noindent
Moreover,  let us denote by $\textit{Spec}'$ the set of clauses obtained from $\textit{Spec}^{\sharp}$ by replacing all occurrences of $f$ by $r_{\textit{prog}}$.
We show that  $M(\textit{OpSem}\cup \textit{Aux}\cup \textit{Dom}) \models \textit{Spec}'$.

\noindent
Let  $S$ be any clause in $\textit{Spec}'$. If $S$ belongs to  $\textit{Aux}$, then $M(\textit{OpSem}\cup \textit{Aux}) \models S$.
Otherwise, $S$ is of the form $r_{\textit{prog}}(X_1,\ldots,X_s,Y) \leftarrow \textit{dom}_r(X_1,\ldots,X_s), \widetilde{B}$ and, by construction,  in $F_{\textit{pcorr}}$
there are two goals

\smallskip
$G_1$: $\textit{false}  \leftarrow Y\!>\!Z, r_{\textit{prog}}(X_1,\ldots,X_s,Z), \widetilde{B}$,
and 

$G_2$: $\textit{false} \leftarrow Y\!<\!Z, r_{\textit{prog}}(X_1,\ldots,X_s,Z), \widetilde{B}$ 

\smallskip
\noindent
such that $\textit{OpSem}\cup \textit{Aux} \cup \{G_1,G_2\}$ is satisfiable.
Then, 

\smallskip
$M(\textit{OpSem}\cup \textit{Aux}) \models \neg\exists (Y\neq Z \wedge r_{\textit{prog}}(X_1,\ldots,X_s,Z) \wedge \widetilde{B})$

\smallskip
\noindent
Since $M(\textit{OpSem}\cup \textit{Dom}) \models r_{\textit{prog}}(X_1,\ldots,X_s,Z) \rightarrow \textit{dom}_r(P_1,\ldots,P_s)$, we also have that

\smallskip
$M(\!\textit{OpSem} \cup \textit{Aux} \cup \textit{Dom})\! \models\! \neg\exists (\!Y\!\neq\! Z \wedge \textit{dom}_r(X_1,\ldots,X_s) \wedge r_{\textit{prog}}(X_1,\ldots,X_s,Z) \wedge \widetilde{B})$

\smallskip
\noindent
From the functionality of $ r_{\textit{prog}}$ it follows that 

\smallskip
$M(\textit{OpSem}\cup \textit{Aux}\cup \textit{Dom}) \models \neg r_{\textit{prog}}(X_1,\ldots,X_s,Y)$ 

\hfill
$ \leftrightarrow (\neg \exists Z. r_{\textit{prog}}(X_1,\ldots,X_s,Y) \vee (r_{\textit{prog}}(X_1,\ldots,X_s,Z) \wedge Y\!\neq\! Z))$

\smallskip
\noindent
 and hence, by using (1), 

\smallskip
$M(\textit{OpSem}\cup \textit{Aux}\cup \textit{Dom}) \models \neg\exists (\textit{dom}_r(X_1,\ldots,X_s) \wedge\neg r_{\textit{prog}}(X_1,\ldots,X_s,Y) \wedge \widetilde{B})$

\smallskip
\noindent
Thus, we have that 

\smallskip
$M(\textit{OpSem}\cup \textit{Aux}\cup \textit{Dom}) \models\forall (\textit{dom}_r(X_1,\ldots,X_s)  \wedge \widetilde{B} \rightarrow r_{\textit{prog}}(X_1,\ldots,X_s,Y))$

\smallskip
\noindent
that is, clause
$S$ is true in $M(\textit{OpSem}\cup \textit{Aux} \cup \textit{Dom})$.
We can conclude that $M(\textit{OpSem}\cup \textit{Aux}\cup \textit{Dom})$ is a model of
$\textit{Spec}'\cup \textit{Dom}$, and since $M(\textit{Spec}'\cup \textit{Dom})$ is the {\it least} model of $\textit{Spec}'\cup \textit{Dom}$, we have that

\smallskip
$M(\textit{Spec}'\cup \textit{Dom}) \subseteq M(\textit{OpSem}\cup \textit{Aux}\cup \textit{Dom})$ \hfill (3)

\smallskip
\noindent
Next we show that,  for all integers $p_1,\ldots,p_s,y$,

\smallskip

$M(\textit{Spec}^{\sharp} \cup \textit{Dom} ) \models f(p_1,\ldots,p_s,y)$  
\ \ iff  \ \
$M(\textit{OpSem}) \models r_{\textit{prog}}(p_1,\ldots,p_s,y)$ \hfill (4)

\smallskip\noindent
{\it Only If Part of} (4).
Suppose  that
$M(\textit{Spec}^{\sharp} \cup \textit{Dom}) \models f(p_1,\ldots,p_s,y)$. Then, by 
construction,

\smallskip

$M(\textit{Spec}'\cup \textit{Dom} ) \models r_{\textit{prog}}(p_1,\ldots,p_s,y)$

\smallskip

\noindent
and hence, by (3),

\smallskip

$M(\textit{OpSem} \cup \textit{Aux}\cup \textit{Dom}) \models r_{\textit{prog}}(p_1,\ldots,p_s,y)$ 

\smallskip

\noindent
Since $r_{\textit{prog}}$ does not depend on predicates in $ \textit{Aux}\cup \textit{Dom}$,

\smallskip

$M(\textit{OpSem}) \models r_{\textit{prog}}(p_1,\ldots,p_s,y)$

\medskip

\noindent
{\it If Part of} (4). 
Suppose that
$M(\textit{OpSem}) \models r_{\textit{prog}}(p_1,\ldots,p_s,y)$.

\noindent
Then, by definition of  $ r_{\textit{prog}}$,

\smallskip
$M(\textit{Dom}) \models {\textit{dom}_r}(p_1,\ldots,p_s)$ \hfill (5)

\noindent
and 

$M(\textit{Spec}) \models {\textit{pre}}(p_1,\ldots,p_s)$  \hfill (6)

\smallskip
\noindent
Thus, by (6) and {Condition~(3.1)}
of Definition \ref{def:spec}, there exists $z$ such that 

\smallskip
$M(\textit{Spec})\models f(p_1,\ldots,p_s,z) $ \hfill (7)

\smallskip
\noindent
By (5) and (7),

\smallskip
$M(\textit{Spec}^{\sharp}\cup \textit{Dom})\models f(p_1,\ldots,p_s,z) $ \hfill (8)

\smallskip
\noindent
By the {\it Only If Part} of (4), 

\smallskip
$M(\textit{OpSem})\models r_{\textit{prog}}(p_1,\ldots,p_s,z)$ 

\smallskip
\noindent
and
by the functionality of $ r_{\textit{prog}}$, $z=y$. Hence, by (8),

\smallskip
$M(\textit{Spec}^{\sharp}\cup \textit{Dom})\models f(p_1,\ldots,p_s,y) $

\medskip
\noindent
Let us now prove partial correctness.
If $M(\textit{Spec}) \models {\textit{pre}}(p_1,\ldots,p_s)$  and 
\textit{prog} terminates, that is, $M(\textit{Dom}) \models \textit{dom}_r(p_1,\ldots,p_s)$,  
then for some integer~$y$, $M(\textit{OpSem}) \models  r_{\textit{prog}}(p_1,\ldots,p_s,y)$.
Thus, by (4),  $M(\textit{Spec}^\sharp \cup \textit{Dom}) \models  f(p_1,\ldots,p_s,y)$
and hence, by (2), $M(\textit{Spec}) \models  f(p_1,\ldots,p_s,y)$.
Suppose that  the postcondition $\psi$ is $f(p_1,\ldots,p_s,z_k)$. Then,
by {Condition~(3.2)} of Definition~\ref{def:spec}, $y=z_k$.

Thus, \mbox{$\{\varphi\} $ {\textit{prog}} $\{\psi\}$}.
\hfill $\Box$


\bigskip

\noindent
{\it Removal of the Interpreter}

\noindent
Here we report the variant of the transformation presented in \cite{De&14c} that
we use in this paper to perform the removal of the interpreter. 
In this transformation we use the function $\textit{Unf\/}(C,A,{\it Cls})$ 
defined as the set of clauses derived by unfolding a clause~$C$ with respect to an atom~$A$
using the set~$\textit{Cls}$ of clauses (see the unfolding rule in Section \ref{subsec:linearization}).

The predicate {\it reach} is defined as follows:

${\it reach}(X,X) \leftarrow$

${\it reach}(X,Z) \leftarrow {\it tr}(X,Y),\ {\it reach}(Y,Z)$

\noindent
where, as mentioned in Section \ref{sec:encoding}, {\it tr} is a (nonrecursive) predicate representing 
one transition step according to the operational semantics of the imperative language.

In order to perform the \textsc{Unfolding} step,
we assume that the atoms occurring in bodies of clauses 
are annotated as either {\em unfoldable} 
or {\em not unfoldable}. 
This annotation 
ensures that any sequence of clauses constructed by unfolding
w.r.t.~unfoldable atoms is finite. 
In particular, the atoms with predicate \textit{initCf\/}, \textit{finalCf\/},
and {\it tr\/} are unfoldable. The atoms of the form 
$\textit{reach}(\textit{cf}_1,\textit{cf}_2)$ are unfoldable if $\textit{cf}_1$
is not associated with a while or goto command. 
Other annotations based on a different analysis of program ${\it OpSem}$ can be used.

	\noindent\hrulefill
	
	\noindent \emph{Input\/}: Program~$\textit{OpSem}$.\\ 
	\noindent \emph{Output\/}: Program $\textit{OpSem}_{\textit{\small{RI}}}$ such that,
for all integers $p_1,\ldots,\!p_s,\!z_k$,
	 
	$r_{\textit{\small prog}}(\!p_1,\ldots,\!p_s,\!z_k)\!\in\! M(\textit{OpSem})$ iff 
$r_{\textit{\small prog}}(\!p_1,\ldots,\!p_s,\!z_k)\!\in\! M(\textit{OpSem}_{\textit{\small{RI}}})$.\\	
\noindent\rule{30mm}{0.1mm}
	
	\noindent \textsc{Initialization}:

\noindent\hspace{.6mm}
$\textit{OpSem}_{\textit{\small{RI}}}:=\emptyset$; ~~$\textit{Defs}:= \emptyset$;

\noindent\hspace{.6mm}
$\textit{InCls}:=\!\{ 
r_{\textit{\small prog}}(\!P_1,\!\ldots\!,\!P_s,\!Z_k\!) 
\!\leftarrow\! \textit{initCf\/}
(\!C_0,\!P_1,\!\ldots\!,\!P_s), \textit{reach}(\!C_0,\!\textit{C}_h), \textit{finalCf\/}(\!C_h,\!Z_k)\!\}$;
	
	
	\noindent \textit{while}~in \textit{InCls} there is a clause~$C$
	which is not a constrained fact
	\textit{do}
	
	\smallskip
	
	\hspace*{3mm}\begin{minipage}{118mm}
		
		\smallskip
		\noindent \textsc{Unfolding}:
		
		
		\noindent $\textit{SpC} :=\textit{Unf}(C,A,\textit{OpSem})$, 
		where \textit{A} is the leftmost atom in the body of~$C$;

		\noindent \textit{while}
		\hangindent=8mm in $\textit{SpC}$ there is a clause $D$
		whose body contains an occurrence of an unfoldable atom \textit{A}
		\textit{do} 
		\\
		$\textit{~~~SpC} := (\textit{SpC}- \{ D\}) \cup\ \textit{Unf}(D,A,\textit{OpSem})$
			
		\noindent \textit{end-while};

		\smallskip
		
		\noindent {\textsc{Definition} \& \textsc{Folding}:}
		
		\noindent
		\makebox[8mm][l]{\textit{while}}\ in $\textit{SpC}$ there is a clause
		$E$  of the form: 
		~~$H\leftarrow e, \textit{reach}(\textit{cf}_1,\textit{cf}_2)$
		
		\noindent  
		\textit{do}
		
		\smallskip
		
		\hspace*{3mm}\begin{minipage}{115mm}
			
			\noindent
			\makebox[4mm][l]{\textit{if}}in $\textit{Defs}$ there is no clause of the form:
			~~$\textit{newp}(V)\leftarrow \textit{reach}(\textit{cf}_1,\textit{cf}_2)$ 
			
			\hspace*{4mm}where {\it V} is the set of variables occurring in $\textit{reach}(\textit{cf}_1,\textit{cf}_2)$

			\noindent
			\textit{then}~~add the clause $N$: $\textit{newp}(V)\leftarrow \textit{reach}(\textit{cf}_1,\textit{cf}_2)$
			to \textit{Defs} and \textit{InCls};
						
			$\textit{SpC} :=(\textit{SpC} -\{E \})\cup \{H\leftarrow e, \textit{newp}(V)\}$
		\end{minipage}
		
		\smallskip
		\noindent
		\textit{end-while};
		
		\smallskip
		
		\noindent $\textit{InCls}:=\textit{InCls}-\{C\}$; ~~
		$\textit{OpSem}_{\textit{\small{RI}}}:=\textit{OpSem}_{\textit{\small{RI}}}\cup \textit{SpC}$;
		
	\end{minipage}
	
	\smallskip
	
	\noindent \textit{end-while};
	
	\vspace*{-1.5mm}
	\noindent\hrulefill\nopagebreak
\vspace{-2mm}
\begin{center} RI: Removal of the Interpreter. \end{center}
\nopagebreak

\noindent
Let us now prove Theorem \ref{thm:RI} stating the relevant properties of the RI transformation.

\medskip
\noindent
{\it The RI transformation terminates.} 
The termination of the \textsc{Unfolding} step is guaranteed by the 
{\it unfoldable} annotations. Indeed, (i) the repeated unfolding of the
unfoldable atoms with predicates \textit{initCf\/}, \textit{finalCf\/}, and {\it tr}, 
always terminates because those atoms have no recursive clauses,
(ii) by the definition of the semantics of the imperative program, 
the repeated unfolding of an atom of the form
$\textit{reach}(\textit{cf}_1,\textit{cf}_2)$ eventually derives
a new $\textit{reach}(\textit{cf}_3,\textit{cf}_4)$ atom where
 $\textit{cf}_3$ is either a final configuration or
a configuration associated with a while or goto command, and in both cases
unfolding terminates.
The termination of the \textsc{Definition} \& \textsc{Folding} step
follows from the fact that {\it SpC} is a finite set of clauses. 

The outer while loop terminates because a finite set of new predicate definitions
of the form $\textit{newp}(V)\leftarrow \textit{reach}(\textit{cf}_1,\textit{cf}_2)$
can be introduced.
Indeed, each configuration {\it cf} is represented as a term 
{\tt cf(LC,E))}, where {\tt LC} is a labeled command and 
{\tt E} is an environment (see Example \ref{ex:fib}). 
An environment is represented as a list of $(v,X)$ pairs where 
$v$ is a variable identifier and $X$ is its value, that is, a logical
variable whose value may be subject to a given constraint.
Considering that: (i) the labeled commands and the variable identifiers
occurring in an imperative 
program are finitely many, and
(ii) predicate definitions of the form
 $\textit{newp}(V)\leftarrow \textit{reach}(\textit{cf}_1,\textit{cf}_2)$
abstract away from the constraints that hold on the logical
variables occurring in $\textit{cf}_1$ and $\textit{cf}_2$,
we can conclude that there are only finitely many such clauses (modulo variable renaming).

\medskip	
\noindent
{\it Point}~1: {\it $\textit{OpSem}_{\textit{\small{RI}}}$ is a set of linear clauses over the integers.}
By construction, every clause in $\textit{OpSem}_{\textit{\small{RI}}}$ is of the 
form $H \leftarrow c, B$, where
(i) $H$ is either $r_{\textit{\small prog}}(\!P_1,\ldots,\!P_s,\!Z_k)$ or $\textit{newp}(V)$, for some new predicate
$\textit{newp}$ and tuple of variables $V$, and
(ii) $B$ is either absent or of the form $\textit{newp}(V)$, for some new predicate
$\textit{newp}$ and tuple of variables $V$.
Thus, every clause is a linear clause over the integers.
	
\medskip	
\noindent
{\it Point}~2: {\it $\textit{OpSem} \cup \textit{Aux}\, \cup F_{\textit{\small{pcorr}}}$ is satisfiable iff 
	$\textit{OpSem}_{\textit{\small{RI}}} \cup \textit{Aux}\, \cup F_{\textit{\small{pcorr}}}$ is satisfiable.}
From the correctness of the unfolding, definition, and folding rules with respect
to the least model semantics of CLP programs \cite{EtG96}, it follows that,
for all integers $p_1,\ldots,\!p_s,\!z_k$,

$r_{\textit{\small prog}}(p_1,\ldots,\!p_s,\!z_k)\!\in\! M(\textit{OpSem})$ iff 
	$r_{\textit{\small prog}}(p_1,\ldots,\!p_s,\!z_k)\!\in\! M(\textit{OpSem}_{\textit{\small{RI}}})$ \hfill $(\dagger1)$ ~~~
	
\noindent
$\textit{OpSem} \cup \textit{Aux}\, \cup F_{\textit{\small{pcorr}}}$ is satisfiable iff 
for every ground instance $G$ of a goal in $F_{\textit{\small{pcorr}}}$, $M(\textit{OpSem} \cup \textit{Aux})\models G$.
Since the only predicate of \textit{OpSem} on which $G$ may depend is $r_{\textit{\small prog}}$, by $(\dagger1)$,
we have that $M(\textit{OpSem} \cup \textit{Aux})\models G$ iff 
$M(\textit{OpSem}_{\textit{\small{RI}}} \cup \textit{Aux})\models G$.
Finally,  $M(\textit{OpSem}_{\textit{\small{RI}}} \cup \textit{Aux})\models G$
for every ground instance $G$ of a goal in $F_{\textit{\small{pcorr}}}$, iff
$\textit{OpSem}_{\textit{\small{RI}}} \cup \textit{Aux}\, \cup F_{\textit{\small{pcorr}}}$ is satisfiable.

\medskip	
\noindent
{\it Point}~3: {\it $\textit{OpSem}\, \cup \textit{Aux}\, \cup F_{\textit{\small{pcorr}}}$ is 
	${\textit{LA}}$-solvable iff $\textit{OpSem}_{\textit{\small{RI}}}\, \cup 
	\textit{Aux}\, \cup F_{\textit{\small{pcorr}}}$ is {\textit{LA}}-solvable.}

Suppose that $\textit{OpSem}\, \cup \textit{Aux}\, \cup F_{\textit{\small{pcorr}}}$ is 
${\textit{LA}}$-solvable, and let $\Sigma$ be an {\it LA}-solution of 
$\textit{OpSem}\, \cup \textit{Aux}\, \cup F_{\textit{\small{pcorr}}}$. 
Now we construct an {\it LA}-solution $\Sigma_{\textit{\small{RI}}}$ of $\textit{OpSem}_{\textit{\small{RI}}}\, \cup 
	\textit{Aux}\, \cup F_{\textit{\small{pcorr}}}$.
To this purpose it is enough to define a symbolic interpretation for
the new predicates introduced by RI.

For any predicate {\it newp} introduced by RI via a clause of the form:

\smallskip

$\textit{newp}(V)\leftarrow \textit{reach}(\textit{cf}_1,\textit{cf}_2)$

\smallskip

\noindent
we define a symbolic interpretation as follows:

\smallskip

$\Sigma_{\textit{\small{RI}}}(\textit{newp}(V)) = \Sigma(\textit{reach}(\textit{cf}_1,\textit{cf}_2))$

\smallskip

\noindent
Moreover, $\Sigma_{\textit{\small{RI}}}$ is identical to $\Sigma$ for the atoms with
predicate occurring in \textit{OpSem}.

Now we have to prove that $\Sigma_{\textit{\small{RI}}}$ is indeed an
{\it LA}-solution of $\textit{OpSem}_{\textit{\small{RI}}}\, \cup 
\textit{Aux}\, \cup F_{\textit{\small{pcorr}}}$. This proof is similar to
the proof of Theorem \ref{thm:LAsolv} (actually, simpler, because RI
introduces new predicates defined by single atoms, while LIN introduces
new predicates defined by conjunctions of atoms), and is omitted.

Vice versa, if $\Sigma_{\textit{\small{RI}}}$ is an {\it LA}-solution
of $\textit{OpSem}_{\textit{\small{RI}}}\, \cup \textit{Aux}\, \cup F_{\textit{\small{pcorr}}}$,
we construct an {\it LA}-solution $\Sigma$ of 
 $\textit{OpSem}\, \cup \textit{Aux}\, \cup F_{\textit{\small{pcorr}}}$
by defining

$\Sigma(\textit{reach}(\textit{cf}_1,\textit{cf}_2)) = \Sigma_{\textit{\small{RI}}}(\textit{newp}(V))$.
\hfill \eop

\bigskip

\noindent
{\it Proof of Theorem \ref{thm:term-corr}} 

\noindent
Let {\it LCls} be a set of linear clauses and \textit{Gls} be a set of nonlinear goals. 
We split the proof of Theorem \ref{thm:term-corr} in three parts:

\noindent
{\it Termination}: The \textrm{linearization} transformation LIN terminates for the input set of clauses $\textit{LCls}\, \cup \textit{Gls}$;

\noindent
{\it Linearity}: The output $\textit{TransfCls}$ of LIN is a set of linear clauses;

\noindent
{\it Equisatisfiability}: $\textit{LCls}\, \cup \textit{Gls}$ is satisfiable iff
	\textit{TransfCls} is satisfiable.

\smallskip

\noindent
({\it Termination}) 
Each \textsc{Unfolding} and \textsc{Definition}\,\&\,\textsc{Folding} step terminates.
Thus, in order to prove the termination of LIN it is enough to show
that the while loop is executed a finite number of times, that is,
a finite number of clauses are added to {\it NLCls}.
We will establish this finiteness property by showing that there exists
an integer $M$ such that every clause added to {\it NLCls} is of the form:

$\textit{newp}(X_1,\ldots,X_t) \leftarrow A_1, \ldots, A_k$ \hfill $(\dagger2)$ ~~~

\noindent
where: (i) $k\leq M$, (ii) for $i=1,\ldots,k$, $A_i$ is of the form
$p(X_1,\ldots,X_m)$, and (iii)~$\{X_1,\ldots,X_t\} \subseteq \textit{vars}(A_1, \ldots, A_k)$.

Indeed, let $M$ be the maximal number of atoms occurring in the body
of a goal in {\it Gls}, to which {\it NLCls} 
is initialized.
Now let us consider a clause $C$ in {\it NLCls} and assume that
in the body of $C$ there are at most $M$ atoms.
The clauses in the set \textit{LCls} used for unfolding $C$ are linear, and hence
in the body of each clause belonging to the set $U(C)$ obtained after the \textsc{Unfolding}
step, there are at most $M$ atoms.
Thus, each clause in $U(C)$
is of the form $H \leftarrow c, A_1, \ldots, A_k$, with $k\leq M$.
Since the body of every new clause introduced by the subsequent
\textsc{Definition}\,\&\,\textsc{Folding} step is obtained 
by dropping the constraint from the body of a clause in $U(C)$,
we have that every clause added to {\it NLCls} is of the form 
 $(\dagger2)$,
with $k\leq M$.
Thus, LIN terminates.

\smallskip
\noindent
({\it Linearity}) \textit{TransfCls} is initialized to the set
\textit{LCls} of linear clauses. Moreover, each clause added to \textit{TransfCls}
is of the form $H \leftarrow c,\, \textit{newp}(X_1,\ldots,X_t)$, and hence
is linear.

\smallskip
\noindent
({\it Equisatisfiability}) 
In order to prove that LIN ensures equisatisfiability, let us 
adapt to our context the basic notions about the unfold/fold 
transformation rules for CLP programs presented in~\cite{EtG96}.

Besides the unfolding rule of Section \ref{subsec:linearization}, 
we also introduce the following {\it definition} and {\it folding} rules.

\smallskip

\noindent
{\it Definition Rule.}  By definition we 
introduce a clause of the form \mbox{$\textit{newp}(X) \leftarrow G$}, 
where \textit{newp} is a new predicate symbol and
$X$ is a tuple of variables occurring in~\textit{G}.

\smallskip

\noindent
{\it Folding Rule.} 
Given a clause $E$: ${H\leftarrow c, G}$ and 
a clause $D$: ${\textit{newp}(X)\leftarrow G}$ introduced by the definition rule.
Suppose that, ${\textit X} = \textit{vars}({\it G}) \cap \textit{vars}(\textit{H,c})$.
Then by folding~$E$ using $D$ we derive $H\leftarrow c, \textit{newp}(X)$.

\smallskip

From a set {\it Cls} of clauses
we can derive a new set $\textit{TransfCls}$ of clauses either by
adding a new clause to {\it Cls} using the definition rule or by: 
(i)~selecting a clause~$C$ in {\it Cls},
(ii)~deriving a new set \textit{TransfC} of clauses using
one or more transformation rules among unfolding and folding, and
(iii)~replacing $C$ by \textit{TransfC} in~{\it Cls}.
We can apply a new sequence of transformation rules
starting from \textit{TransfCls} and iterate this process at will.

The following theorem is an immediate consequence of the correctness results 
for the unfold/fold transformation rules of CLP programs~\cite{EtG96}. 

\begin{theorem}[{\textit Correctness of the Transformation Rules}]
\label{thm:corr_rules}
Let the set $\textit{TransfCls}$ be derived from {\it Cls} by a sequence
of applications of the unfolding, definition and folding transformation rules. 
Suppose that every clause introduced by the definition
rule is unfolded at least once in this sequence.
Then, {\it Cls} is satisfiable \mbox{iff}
{\it TransfCls} is satisfiable.
\end{theorem}

Now, equisatisfiability easily follows from 
Theorem \ref{thm:corr_rules}. Indeed, the \textsc{Unfolding} and
\textsc{Definition \& Folding} steps of LIN
are applications of the unfolding, definition, and folding rules
(strictly speaking, the rewriting performed after unfolding is not included 
among the transformation rules, but obviously preserves all {\it LA}-models). 
Moreover, every clause introduced during the \textsc{Definition \& Folding} step
is added to {\it NCls} and unfolded in a subsequent step of the transformation.
Thus, the hypotheses of Theorem \ref{thm:corr_rules} are fulfilled, and hence
we have that $\textit{LCls}\, \cup \textit{Gls}$ is satisfiable iff
\textit{TransfCls} is satisfiable. \hfill $\Box$

\bigskip

\noindent
{\it Linearized clauses for Fibonacci.} 

\noindent
The set of {\it linear} constrained Horn clauses obtained after applying LIN
is made out of clauses {\tts E1}, {\tts E2}, 
{\tts E3}, and {\tts C3}, together with the following clauses:
\vspace{1.5mm}

\smallskip
{\small{\tts{

\noindent
new1(N1,U,V,U,N2,U,N3,U):-\,N1=<0,\,N2=<0,\,N3=<0.

\noindent
new1(N1,U,V,U,N2,U,N3,F3):-\,N1=<0,\,N2=<0,\,N4=N3-1,\,W=U+V,\,N3>=1,
             new2(N4,W,U,F3).

\noindent             
new1(N1,U,V,U,N2,F2,N3,U):-\,N1=<0,\,N4=N2-1,\,W=U+V,\,N2>=1,\,N3=<0,
             new2(N4,W,U,F2).
 
\noindent            
new1(N1,U,V,U,N2,F2,N3,F3):-\,N1=<0,\,N4=N2-1,\,N2>=1,\,N5=N3-1,\,N3>=1,\par\hspace{12mm}  
             new3(N4,W,U,F2,N5,F3).

\noindent             
new1(N1,U,V,F1,N2,U,N3,U):-\,N4=N1-1,\,W=U+V,\,N1>=1,\,N2=<0,\,N3=<0,
             new2(N4,W,U,F1).
 
\noindent            
new1(N1,U,V,F1,N2,U,N3,F3):-\,N4=N1-1,\,N1>=1,\,N2=<0,\,N5=N3-1,\,W=U+V,\,N3>=1,\nopagebreak\par\hspace{12mm}  
             new3(N4,W,U,F1,N5,F3).

\noindent             
new1(N1,U,V,F1,N2,F2,N3,U):-\,N4=N1-1,\,N1>=1,\,N5=N2-1,\,W=U+V,\,N2>=1,\,N3=<0,\nopagebreak\par\hspace{12mm}   
             new3(N4,W,U,F1,N5,F2).

\noindent             
new1(N1,U,V,F1,N2,F2,N3,F3):-\,N4=N1-1,\,N1>=1,\,N5=N2-1,\,N2>=1,\,N6=N3-1,\,W=U+V,
\nopagebreak\par\hspace{12mm} N3>=1,\,new1(N4,W,U,F1,N5,F2,N6,F3).
 
\noindent            
new2(N,U,V,U):-\,N=<0.

\noindent
new2(N,U,V,F):-\,N2=N-1,\,W=U+V,\,N>=1,\,new2(N2,W,U,F).

\noindent
new3(N1,U,V,U,N2,U):-\,N1=<0,\,N2=<0.

\noindent
new3(N1,U,V,U,N2,F2):-\,N1=<0,\,N3=N2-1,\,W=U+V,\,N2>=1,\,new2(N3,W,U,F2).

\noindent
new3(N1,U,V,F1,N2,F2):-\,N3=N1-1,\,N1>=1,\,N4=N2-1,\,W=U+V,\,N2>=1,\par\hspace{12mm} 
             new3(N3,W,U,F1,N4,F2).
             
\noindent
new3(N1,U,V,F1,N2,U):-\,N3=N1-1,\,W=U+V,\,N1>=1,\,N2=<0,\,new2(N3,W,U,F1).

}}} 
\normalsize

\medskip

\noindent
{\it Proof of Theorem \ref{thm:LAsolv} $($Monotonicity with respect to {\it LA}-Solvability$)$.}

\noindent
Suppose that the set $\textit{LCls}\cup \textit{Gls}$ of constrained Horn clauses
is {\it LA}-solvable, and let $\textit{TransfCls}$ be obtained by applying
LIN to $\textit{LCls}\cup \textit{Gls}$. 
Let $\Sigma$ be an {\it LA}-solution of $\textit{LCls}\cup \textit{Gls}$. We now construct an {\it LA}-solution of $\textit{TransfCls}$.
For any predicate {\it newp} introduced by LIN via a clause of the form:

$\textit{newp}(X_1,\ldots,X_t) \leftarrow A_1, \ldots, A_k$

\noindent
we define a symbolic interpretation $\Sigma'$ as follows:

$\Sigma'(\textit{newp}(X_1,\ldots,X_t)) = \Sigma(A_1) \wedge \ldots \wedge \Sigma(A_k)$

\noindent
Now, we are left with the task of proving that $\Sigma'$ is indeed an
{\it LA}-solution of $\textit{TransfCls}$.
The clauses in $\textit{TransfCls}$ are either of the form

$\textit{false} \leftarrow c, \textit{newq}(X_1,\ldots,X_u)$

\noindent
or of the form

$\textit{newp}(X_1,\ldots,X_t) \leftarrow c, \textit{newq}(X_1,\ldots,X_u)$

\noindent
where \textit{newp} and \textit{newq} are
 predicates  introduced by LIN. We will only consider the more difficult case where
the conclusion is not \textit{false}.

The clause $\textit{newp}(X_1,\ldots,X_t) \leftarrow c, \textit{newq}(X_1,\ldots,X_u)$
has been derived (see the {\rm linearization} transformation LIN in Figure \ref{fig:Lin})
in the following two steps.

\noindent
(Step i) Unfolding  $\textit{newp}(X_1,\ldots,X_t) \leftarrow A_1, \ldots, A_k$
w.r.t. all atoms in its body using $k$ clauses in \textit{LCls}:

$A_1 \leftarrow c_1, B_1 \ \ \ldots \ \  A_k\leftarrow c_k, B_k$

\noindent
where some of the $B_i$'s can be the \textit{true} and $c \equiv c_1,\ldots,c_k$, thereby deriving

$\textit{newp}(X_1,\ldots,X_t) \leftarrow c_1, \ldots, c_k, B_1, \ldots, B_k$

\noindent
(Without loss of generality we assume that the atoms in the body
of the clauses are equal to, instead of {\it unifiable} with,
the heads of the clauses in \textit{LCls}. )

\noindent
(Step ii) Folding $\textit{newp}(X_1,\ldots,X_t) \leftarrow c_1, \ldots, c_k, B_1, \ldots, B_k$
using a clause of the form:

$\textit{newq}(X_1,\ldots,X_u) \leftarrow B_1, \ldots, B_k$

\noindent
Thus, for $\textit{newq}(X_1,\ldots,X_u)) $ we have the following symbolic interpretation:

$\Sigma'(\textit{newq}(X_1,\ldots,X_u)) = \Sigma(B_1) \wedge \ldots \wedge \Sigma(B_k)$

\medskip
\noindent
To prove that $\Sigma'$ is an {\it LA}-solution of $\textit{TransfCls}$, we have to show that

$\textit{LA} \models \forall (c \wedge \Sigma'(\textit{newq}(X_1,\ldots,X_u)) \rightarrow \Sigma'(\textit{newp}(X_1,\ldots,X_t)))$

\noindent
Assume that

$\textit{LA} \models c \wedge \Sigma'(\textit{newq}(X_1,\ldots,X_u)) $

\noindent
Then, by definition of $\Sigma'$,

$\textit{LA} \models c \wedge \Sigma(B_1) \wedge \ldots \wedge \Sigma(B_k) $

\noindent
Since $\Sigma$ is an {\it LA}-solution of \textit{LCls}, we have that:

${\it LA} \models \forall (c_1 \wedge \Sigma(B_1) \rightarrow \Sigma(A_1))  \ \ \ldots \ \  \textit{LA} \models \forall (c_k \wedge \Sigma(B_k) \rightarrow \Sigma(A_k)) $

\noindent
and hence

$\textit{LA} \models \Sigma(A_1) \wedge \ldots \wedge \Sigma(A_k) $

\noindent
Thus, by definition of $\Sigma'$,

$\textit{LA} \models \Sigma'(\textit{newp}(X_1,\ldots,X_t)).
$ \hfill \eop

\end{document}